\documentclass{elsart}

\usepackage{amssymb}

\usepackage{graphicx}
\usepackage{dcolumn}
\usepackage{bm}

\begin{document}

\begin{frontmatter}


\title{Effects of the energy error distribution of fluorescence telescopes on  
the UHECR energy spectrum} 

\author[if]{Washington Carvalho Jr.,} \ead{carvajr@if.usp.br}
\author[if]{Ivone F.M. Albuquerque \corauthref{cor1}}
\ead{ifreire@if.usp.br}
\author[iag]{Vitor de Souza} \ead{vitor@astro.iag.usp.br}
\address[if]{Instituto de F\'{\i}sica, Universidade de S\~ao Paulo,
  Brasil}
\address[iag]{Instituto de Astronomia, Geof\'isica e Ci\^encias
  Atmosf\'ericas \\ Universidade de S\~ao Paulo, Brasil}
\corauth[cor1]{Tel.: +55 11 3091 6862}

\begin{abstract}
The measurement of the ultra high energy cosmic ray (UHECR) spectrum
is strongly affected by uncertainties on the reconstructed energy.
The determination of the presence or absence of the GZK cutoff and its 
position in the energy spectrum depends not only on high statistics but 
also on the shape of the energy error distribution. Here 
we determine the energy error
distribution for fluorescence telescopes, based on a Monte Carlo simulation.
The HiRes and Auger fluorescence telescopes are simulated in detail.
We analyze the UHECR spectrum convolved with this energy error distribution.
We compare this spectrum with one convolved with a lognormal
error distribution as well as with a Gaussian error
distribution. We show that the energy error distribution for fluorescence 
detectors can not be represented by these known distributions. We conclude 
that the convolved energy spectrum will be smeared but
not enough to affect the GZK cutoff detection. This conclusion stands
for both HiRes and Auger fluorescence telescopes. This result differs from the
effect of the energy error distribution obtained with ground detectors and
reinforces the importance of the fluorescence energy measurement. We
also investigate the effect of possible fluorescence yield measurement
errors in the energy spectrum.
\end{abstract}

\begin{keyword}
  cosmic rays \sep energy spectrum \sep fluorescence telescopes
  \sep energy reconstruction error \PACS 96.40-z,96.40.Pq,96.40.De 
\end{keyword}

\end{frontmatter}

\section{\label{introduction} Introduction}

A sharp steepening of the cosmic ray energy spectrum  is expected
around $5 \times 10^{19}$~eV. This cutoff is due to the energy loss of
ultra high energy cosmic rays (UHECR\footnote{cosmic rays with
energies  above $\sim 5 \times 10^{19}$~eV.}) when traveling through
the Cosmic Microwave Background Radiation  (CMB). Greisen \cite{bib:g}
and Zatsepin and Kuzmin \cite{bib:zk,bib:zk2} showed that the main
energy loss mechanism of UHECRs interactions with the CMB is the
photo production of pions. In order to reach the Earth, UHECR should be
produced within $\sim 100$ Mpc radius unless it is a nonstandard
particle \cite{bib:review:gunter,bib:uhe1,bib:uhe2}. As no known
powerful source \cite{bib:som} is located within this range, the
spectrum is expected  to steepen. However, detection of events with
energies above $10^{20}$~eV
\cite{bib:hires:espectro:2,bib:agasa:spectrum} questions the existence
of the steepening of the spectrum and is known as the GZK puzzle. One
of the keys to solve this puzzle is the determination of the UHECR
energy spectrum.

The observation or the absence of the cutoff in the spectrum have different 
consequences. 
The absence will most likely be explained by new physics predicted by models beyond 
the standard model of
particle physics. It is hard to understand the production mechanism and
the composition of the most energetic end of the spectrum
with our current knowledge of astroparticle physics.
If the cutoff is present the understanding of the spectrum will
not necessarily involve new physics but the high end of the energy
spectrum is not yet fully understood.

The recent experimental results from HiRes \cite{bib:hires:espectro:2}
and AGASA \cite{bib:agasa:spectrum} are not conclusive about the
presence of the GZK cutoff. The Pierre Auger Observatory
\cite{bib:auger:nim} is already active and has the capability of
measuring the UHECR spectrum with a statistically significant data set
to determine the presence or absence of the GZK cutoff.

Within this scenario, it is important to assure that the features of
the energy spectrum can be well measured despite of fluctuations that
are intrinsic to the shower development and to energy errors due to
detection  techniques and reconstruction procedures. Recently, it has
been shown \cite{bib:ivone:smoot} that the energy error distribution
(from now on EED) due to the detection of  cosmic rays at ground
level resembles a lognormal distribution. If the standard deviation of
this distribution is too large it will smear the UHECR spectrum in a
way that the GZK cutoff might not be seen. 
The effect of various kinds
of EEDs on the energy spectrum is also done in reference
\cite{bib:tata:escobar}. It has also been argued that the discrepancy
between AGASA \cite{bib:agasa:spectrum} and HiRes
\cite{bib:hires:espectro:2} suggests the presence of systematic errors
in the energy reconstruction \cite{bib:olinto:sistematico}.

In this analysis, we determine the EED for fluorescence telescopes.
We then analyze the UHECR spectrum convolved with this EED. This
convolved spectrum is compared with two different convolutions: one
using a lognormal and another using a Gaussian EED. We show that these
known distributions do not represent well the fluorescence EED.  We
also show that not only the average energy error varies with energy
but also the shape of the EED is energy dependent.

Fluorescence telescopes play an important role in the cosmic ray
energy determination due to its calorimetric measurement. Our
determination of the fluorescence EED includes missing energy
corrections and estimated errors for fluorescence yield
measurements. We determine the EED for two  limiting  cases. First, we
consider only the intrinsic features of air showers and the EED that
arises due to its natural fluctuations. The effect of this EED in the
UHECR spectrum represents the minimum modification that will be
obtained through longitudinal development measurements.

Secondly, a full simulation of the shower detection and reconstruction
procedure is done for fluorescence telescopes and a realistic
smoothing of the  energy spectrum is determined. This analysis has two
steps: the EED is first determined with no   error over the
fluorescence yield simulation and later this error is included in an
arbitrary way. We show that in both cases the EED produces a
noticeable smoothing in the energy spectrum. The simulation with no
fluorescence yield error however does not smear the GZK cutoff
significantly. When these errors are  introduced there is a
significant change in the number of events around the GZK cutoff
position of the spectrum although not big enough to erase the
cutoff. In both cases there is a significant shift on the flux
weighted by a third power of energy.

This article is organized in the following way: in section
\ref{sec:simulation} we describe the shower and telescope simulations,
in section \ref{sec:reconstruction} we determine the EED as a function
of energy for  different telescope configurations and analysis
procedures, in section \ref{sec:spectrum} we analyze the influence of
the EED on the energy spectrum reconstruction and in the last section
we discuss reconstruction procedures, illustrate the differences
between telescope configurations and describe our conclusions.

\section{\label{sec:simulation} Shower and telescope simulation}

As our goal is to determine the effect of  energy measurement errors
on the reconstructed UHECR energy spectrum we determine the EED in two
ways: first, we only consider the errors that arise from natural
shower fluctuations \cite{bib:ivone:smoot} and then we simulate the
EED from fluorescence telescope measurements.

Our Monte Carlo air shower simulation was performed using the CORSIKA
package \cite{bib:corsika}. Hadronic interactions were simulated using
QGSJET01 \cite{bib:qgsjet}. Four sets of 2000 proton induced showers
were simulated. Each set had a different primary energy which ranged
from  $10^{19}$ to $10^{20.5}$ eV with the index varying in steps of
$0.5$.  The thinning factor was set to $10^{-5}$ with a maximum weight
factor of  $10^{6}$. The energy thresholds were set to 0.1 MeV for
electrons and photons; 0.3 GeV for hadrons and 0.7 GeV for muons. The
longitudinal particle and energy deposit profile were sampled in steps
of 5  $g/cm^2$. Each set of 2000 showers were generated at fixed
zenith angle of  60 degrees. Each shower was used several times by
drawing a different zenith angle and  shower core position in relation
to the telescope.

Fluorescence telescopes were simulated according to the general
procedure described in \cite{bib:espectro}. The simulation program was
adapted to produce fluorescence photons from the energy deposited by
the shower at each atmospheric depth.  Photons were generated taking
into account the energy deposition as a function of shower
depth~\cite{bib:kakimoto}, and atmospheric pressure
dependence~\cite{bib:USAtm}. 
These photons were then propagated from the emission point to the
telescope. The attenuation due to Rayleigh (molecular) and Mie
(aerosol) scattering  was taken into account. \v{C}erenkov light is
not taken into account.

Characteristics of the fluorescence detection were simulated in
detail.  The effective collection area, mirror reflectivity,  filter
transmission and phototube quantum efficiency were included in the
telescope efficiency.  We assumed a 20\% telescope efficiency for both
Auger and HiRes-II. The telescope aperture, size of pixels in the
camera, electronic noise and simplified trigger conditions were also
simulated and are defined in section~\ref{subsec:tel}.

The Photo Multiplier Tubes (PMT) were then simulated. The
number of  photons per PMT was determined where both energy threshold
and noise \cite{bib:sim:fd} were taken into account.  The energy
threshold was set as four times the background.  The background is
determined from stable sources of light that are present during data
taking. It yields approximately 40 
photons/m$^2$deg$^2\mu$s~\cite{bib:auger:design}.

\section{\label{sec:reconstruction} Energy reconstruction}

When charged particles transverse the atmosphere  the rotational modes
of nitrogen molecules are excited and fluorescence photons are
emitted. Fluorescence telescopes detect these photons as a function of
the atmospheric depth \cite{bib:review:sokol}.

The number of fluorescence photons produced by an electron traveling
in air is proportional to the electron energy
deposition~\cite{bib:kakimoto,bib:nagano:2}. As the particle's energy as
a function 
of its energy loss is known, the calorimetric energy of the shower is
determined by the number of detected photons. The integration of the
detected fluorescence light over the full shower path determines the
total calorimetric energy of the shower.  This energy plus the energy
of particles that do not generate fluorescence light (denominated by
missing energy) determines the total primary energy of the shower.

The missing energy is carried mainly by high energy muons which lose
most of their energy to the ground and
neutrinos. At ultra high  energies, the correction to account for this
energy \cite{bib:barbosa,bib:song} is around 10\% and has small
deviations for different primary particles and shower
inclinations. In our simulation we correct for the missing energy
according to \cite{bib:barbosa}.

The determination of the shower longitudinal development requires a
fit to the data.  Not only the data is taken in discrete steps as the
longitudinal development of the atmospheric shower is not always fully 
detected by the telescope. For the full evaluation of the shower
longitudinal  development, we use the Gaisser-Hillas function
\cite{bib:gaisser:hillas}.

We first reconstructed the energy taking into account only natural
shower  fluctuations, that is, fluctuations that are intrinsic to the
cascade of particles that constitute the shower
\cite{bib:ivone:smoot}. We fitted  the longitudinal energy deposition
profile simulated by CORSIKA and sampled in 5 g/cm$^2$ with a six
parameters ($E^{dep}_{max}$, $X_{max}$, $X_0$, $a$, $b$ and $c$)
Gaisser-Hillas function:

\begin{equation}
E^{dep}(X) = E^{dep}_{max} \left( \frac{X - X_0}{X_{max} - X_0} \right
) ^{\frac{(X_{max}-X)}{\lambda}} \exp{ \frac{X_{max} -X}{\lambda}}
\end{equation}

where $\lambda = a + bX + cX^2$; $E^{dep}_{max}$ is the energy
deposition at the depth $X_{max}$ where the shower has its maximum
longitudinal size.

The energy of the primary particle was determined by the integration
of the fitted Gaisser-Hillas function plus missing energy corrections.

\begin{figure}[]
\centerline{\includegraphics[width=13cm]{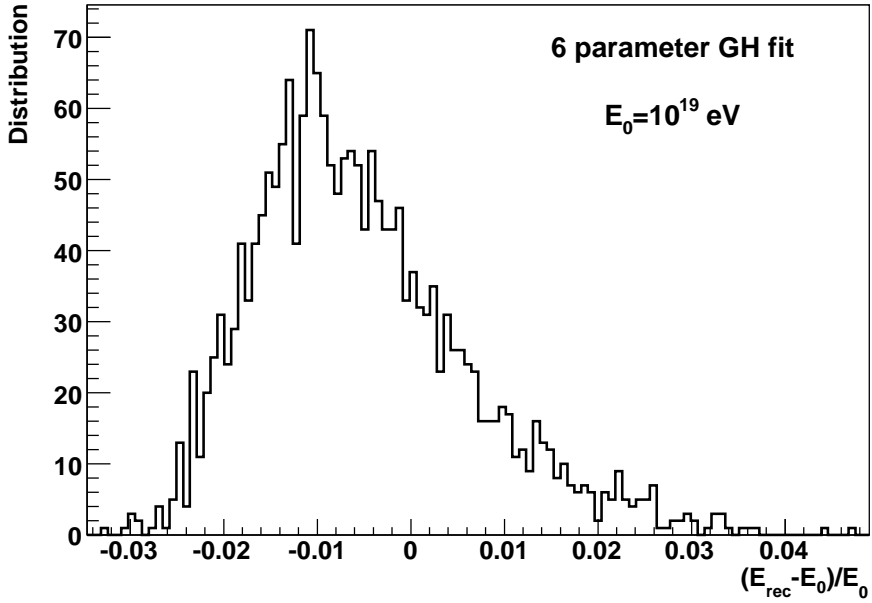}}
\caption{Energy error distribution as reconstructed from 2000
proton $10^{19}$~eV showers. At this level, there are three
main error contributions: a) the  intrinsic fluctuation of the
showers, b) fitting errors of a Gaisser-Hillas function and c) missing
energy corrections.}
\label{fig:erro:corsika}
\end{figure}

Figure \ref{fig:erro:corsika} shows the EED from 2000 simulated
$10^{19}$~eV proton showers generated at a  zenith angle of 60
degrees.  At this level, there are three main contributions to the
reconstructed energy error: a) air shower intrinsic fluctuations, b)
fitting errors and  c) missing energy corrections.  The error
distribution is very narrow, illustrating the effectiveness of the
calorimetric procedure. The maximum energy error is of the order of
4\%.

The influence of this EED in the energy spectrum is minimum and will
be discussed in section \ref{sec:spectrum}.

\subsection{\label{subsec:tel} HiRes-II and Auger Telescopes}

The HiRes-II telescope covers $360^\circ$ in azimuth and  an angle
from $3^\circ$ to $31^\circ$ in elevation with a $1^\circ$ pixel size
camera \cite{bib:hires,bib:jose:tese}. It has  a 5.1~m$^2$
aperture~\cite{bib:hires:espectro:2} and is located at an altitude of
1597~m above sea level. We assume a 20\% telescope efficiency.

The Pierre Auger Observatory uses both ground array and fluorescence
techniques \cite{bib:auger:nim}. The energy spectrum in this
experiment \cite{bib:auger:espectro} is determined from events
detected by its ground array calibrated by the energy measured by its
fluorescence detectors.  This calibration assumes a linear relation
between the signal measured at 1000 meters (S1000) from the shower axis and
the energy measured by the fluorescence telescope. The final EED that
folds into the Auger spectrum is therefore a combination of the
fluorescence telescope error with the ground array energy calibration.

Since the relation between S1000 and energy determined by the
fluorescence telescopes must be fitted, systematic and statistical
errors of both detector 
influences the final EED shape. However, it is clear that the energy scale
of the Auger spectrum is dominated by the fluorescence
telescopes systematic error and that the EED width is most influenced
by the ground detector statistical error.  

In this analysis we took into account the energy errors due to the
fluorescence telescope. One of the four  eyes used by the Auger
Collaboration covering  an angle from $2^\circ$ to $32^\circ$ in
elevation with a  $1.5^\circ$ pixel size camera \cite{bib:sim:fd} was
simulated. The telescope has 10~m$^2$ aperture and is located at an
altitude of 1400~m above sea level. A 20\%  telescope efficiency was
assumed.

\begin{figure}[]
\centerline{\includegraphics[width=13cm]{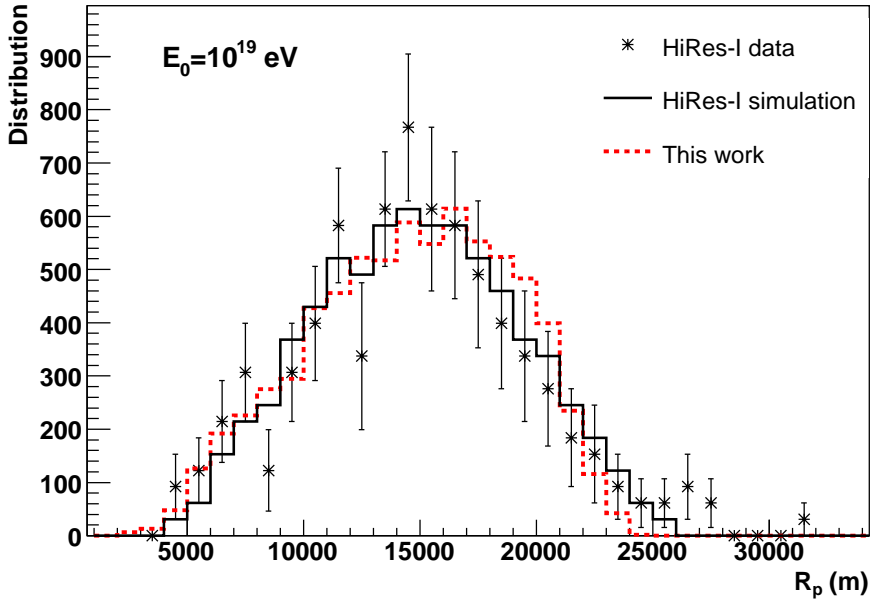}}
\caption{Impact parameter for $10^{19}$ eV proton showers as
  measured ($\ast$) and simulated (black solid line) by the HiRes-I
  Collaboration \cite{bib:hires:espectro} and according to our simulation
  (red dashed line).}
\label{fig:rp}
\end{figure}

In order to test our simulation we determined the impact parameter
from our fluorescence simulation\footnote{for this test we defined our
simulation parameters as in HiRes-I \cite{bib:hires:espectro}.} and
compared it with HiRes-I results
\cite{bib:hires:espectro}. Figure~\ref{fig:rp} shows a good agreement
between our simulation and the HiRes Collaboration analysis.

Using the simulated signal in each PMT of the telescope we
reconstructed the total energy of the shower.  In this procedure, a
$5^\circ$ Gaussian error  was folded into the axis direction within
the shower-detector plane  in order to account for errors in the
reconstruction of the shower axis.  The sequence of hit PMTs was then
transformed back into energy deposited in the atmosphere per path
interval.  Attenuation effects were taken into account using the new
reconstructed shower direction.

The reconstructed energy deposited as a function of atmospheric depth
was fit by a Gaisser-Hillas with two parameters ($E^{dep}_{max}$ and
$X_{max}$)~\footnote{$\lambda$ was set to  70~g/cm$^2$.}. The reduced number of
parameters is due to the reduced number of points measured by the
telescope in comparison to the full longitudinal profile  fit of
section~\ref{sec:reconstruction}. The primary energy was then
determined by adding the missing energy correction \cite{bib:barbosa}
to the integration of the Gaisser-Hillas function.

Quality cuts applied in the UHECR energy spectrum analysis by Auger
\cite{bib:auger:espectro} and HiRes \cite{bib:hires:espectro} were
included in our simulations. They are shown in Table
\ref{tab:cuts}. These cuts are applied in order to clean the data
sample from events which do not allow a good energy
reconstruction. Most of them have only few longitudinal data
points or the $X_{max}$ is not visible.

\begin{figure}[]
\centerline{\includegraphics[width=13cm]{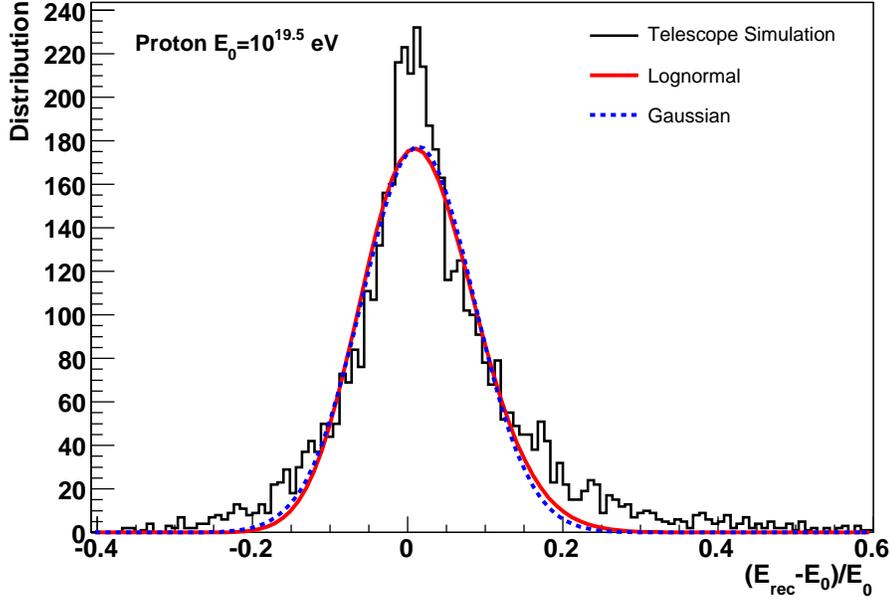}}
\caption{Energy error distribution from simulated fluorescence energy
reconstruction using HiRes-II parameters. A Gaussian with standard
deviation  of 0.07 and lognormal fit with standard deviation of
$\log_{10} {\rm E}$ of 0.03 are superimposed for reference. Both
fits do not include the tails of the distribution.  }
\label{fig:erro:rec}
\end{figure}

Figure~\ref{fig:erro:rec} shows the EED for $10^{19.5}$~eV proton
showers after our simulation of the HiRes-II telescope, reconstruction
procedure and quality cuts. As expected this EED is much wider than
the one shown in Figure~\ref{fig:erro:corsika} with an asymmetric tail
to higher energies. The instrumental detection and energy
reconstruction add sources of errors such as the shower axis geometry
reconstruction, PMT noise, number of hit PMTs and telescope field of
view. These  uncertainties result in a more scarce sampling of the
longitudinal profile.  The contribution of each of these errors to the
overall energy reconstruction is quite difficult to disentangle.

For comparison we fitted the EED shown in  Figure~\ref{fig:erro:rec}
with a Gaussian and a lognormal curve.  The Gaussian standard
deviation ($\sigma$) is 0.07 and the lognormal standard  deviation of
$\log_{10} {\rm E}$ is 0.03 (from here on we will use this  definition
of the lognormal standard deviation). It is clear that neither of
these curves represent well the fluorescence EED.

Figure~\ref{fig:erro:edep} shows the EED for $10^{19}$ and
$10^{20}$~eV  proton showers after our simulation of both HiRes-II and
Auger fluorescence telescopes, including energy reconstruction and
quality cuts.   It can be seen that the EED's shape is different for
each energy. In section~\ref{sec:eedeng} we discuss this energy
dependence on the UHECR spectrum reconstruction.

It is also clear that the asymmetric tail also varies with energy. In
order to better represent this aspect we calculated the skewness of
the EED as a function of energy. Figure~\ref{fig:skewness} shows the
skewness of the EEDs and Table~\ref{tab:en:erro} shows the EED's
average and RMS as a function of energy according to our simulation
of the HiRes-II and Auger telescopes.

\section{\label{sec:spectrum}UHECR Energy Spectrum}

The UHECR energy spectrum at the Earth was determined following the
analysis described in \cite{bib:ivone:smoot}.  Sources were
isotropically distributed at different redshifts and produced a power
law injection spectrum given by:

\begin{equation}
F(E) = kE^{-\alpha} \exp \left ( - \frac{E}{E_{max}} \right )
\label{eq:flux}
\end{equation}

where $E$ is the cosmic ray energy, $k$ is a normalization factor,
$\alpha$ is the spectral index and $E_{max}$ is the maximum cosmic ray
energy at the source. We set $\alpha = 2.6$, $E_{max} = 10^{21}$ eV
and normalize the flux at $10^{19}$~eV as measured by
HiRes-II~\cite{bib:hires:espectro}. Our conclusions are independent
of $\alpha$ \cite{bib:ivone:smoot} and $E_{max}$ is large enough to
not interfere with our results.

Energy losses due to pair production, expansion of the universe and
photo-pion production were included in the propagation of  particles
through the CMB \cite{bib:ivone:smoot}. The energy spectrum at Earth
reproduces the expected GZK cutoff.

We took this UHECR energy spectrum as the true spectrum and convolved
it with the EEDs determined from our simulation of fluorescence
telescopes (described in section \ref{sec:reconstruction}). As we did
not find any good parameterization for the EED as a function of
energy, we defined a Monte Carlo simulation procedure for the
convolution.  The UHECR flux was convoluted in the following way:

\begin{equation}
F'(E) = \int_{0}^{\infty} F(E') P(E',E) dE'
\end{equation}

where F is given by Equation~\ref{eq:flux} and $P(E',E)$ is the
probability that a cosmic ray with energy $E$ has its energy
reconstructed as $E'$.  The energy E' was randomly selected from the
fluorescence EED.

To take into account the EED energy dependency, the fluorescence EED
determined  from $10^{19}$ eV showers was used to convolve the energy
spectrum from below $10^{19}$ to $10^{19.25}$ eV; the EED from $10^{19.5}$
eV  showers to convolve it from $10^{19.25}$ to $10^{19.75}$ eV and
the EED from $10^{20}$ eV showers to convolve the spectrum range above
$10^{19.75}$. No unusual feature appeared in the spectrum due to the
abrupt change of EEDs used in the convolution. In
section~\ref{sec:eedeng} we will discuss the EED energy dependence on
the UHECR  spectrum reconstruction.

Figure \ref{fig:spectrum} shows the UHECR convolved spectrum.  The
solid thin black line represents the true spectrum. The solid thick
red line is the true spectrum convolved with the fluorescence EED from
our simulation of the HiRes-II energy reconstruction. The convolution
with the fluorescence EED determined by our simulation of the Auger
telescope gives a very similar result. The dashed-doted magenta curve
represents the Gaussian (with $\sigma = 0.1{\rm E}$) and the doted
blue curve represents lognormal (with a standard deviation $\log_{10}
{\rm E}$ equal to 0.1) convolutions. The energy spectrum convolved
with an EED determined only from intrinsic shower fluctuations (see
section~\ref{sec:reconstruction}) is very close to the true spectrum
and is not shown.

In order to understand the EED smearing of the spectrum we determined
the excess of events of the convolved spectrum.  Figure \ref{fig:diff}
shows the percentage excess of events determined from a spectrum
convolved with different EEDs compared to the number of events above
$10^{19}$ eV determined from the GZK theoretical prediction (our true
spectrum).  As can be seen, the excess of events is still significant
around the expected GZK energy. Although fluorescence measurements
errors will not erase the GZK cutoff from the spectrum they might
shift its position.

The convolutions with the known Gaussian and lognormal EEDs as well as
the excess of events shown in Figure~\ref{fig:diff} are for  $\sigma =
0.1 {\rm E}$ for the Gaussian and lognormal respectively. Although
these are not representative of the simulated fluorescence EED we
chose standard deviations for these fits close to the ones shown in
Figure~\ref{fig:erro:rec}. Larger lognormal  deviations
\cite{bib:ivone:smoot} will smear the spectrum in a way that the GZK
cutoff will not be seen.

\subsection{\label{sec:eedeng} EED energy dependence effect on UHECR spectrum}

The EED energy dependence can be seen in Figures~\ref{fig:erro:edep}
and~\ref{fig:skewness}.
In order to analyze the effect of this energy dependence on the UHECR
spectrum, we convolved the full spectrum with a EED determined from 
showers generated with the same primary energy.

Figure~\ref{fig:spec:edep} shows the spectrum convolved with 
EEDs from showers with $10^{19}$ and $10^{20}$ eV,
as well as the spectrum convolved with all EEDs as
described in section~\ref{sec:spectrum}. 
It illustrates the difference in the analysis of the energy
spectrum when the EED is evaluated at a single energy from the one
when the EED energy dependence is taken into account.

It can be seen that the spectrum convolved with the $10^{19}$~eV EED 
falls on top of the spectrum convolved with an energy dependent EED.
If however the convolution is done with the $10^{20}$~eV EED, 
the spectrum will differ from the one
in which the EED energy dependence was taken into account.

As the UHECR is a steep falling spectrum the influence between
different EEDs in the spectrum convolution is as expected. As
the flux at $10^{19}$~eV is larger than at higher energies
its influence on the spectrum convolution is also larger.  The EEDs
determined from higher energy showers, have smaller weight in the
convolution and the UHECR spectrum will be significantly different than when
the EED energy dependence is taken into account.

Another important aspect seen from the spectrum convolution with the 
$10^{20}$~eV EED is the effect of the EED shape. Even
small differences in the shape of 
EEDs as those shown in figure \ref{fig:erro:edep} can
significantly modify the energy spectrum (figure \ref{fig:spec:edep}). 
The curve representing the spectrum convolution with an $10^{20}$~eV EED
shows this effect.

\subsection{\label{sec:fy} Uncertainties on the Fluorescence Yield}

The fluorescence yield (FY) of particles in air is an important
parameter in the energy reconstruction procedure. The proportionality
between the energy deposited by an electron and the number of photons
produced per traveled distance is based on a few
measurements~\cite{bib:kakimoto,bib:nagano:2}. Recently,
some new experiments have been proposed in order to measure the
fluorescence yield in a wider energy range and different conditions
including air pressure, composition and temperature
\cite{bib:airfly:2,bib:airlight,bib:lefeuvre,bib:fy:campinas,bib:arqueros,bib:flash}.

In order to analyze the  effect of possible errors in the FY
measurements in the determination of the spectrum we assumed an
arbitrary FY systematic error in our simulation.  This systematic
error was introduced when the energy deposited in the atmosphere was
transformed into fluorescence photons.  We introduced a 10, 30 or 50\%
error in this transformation, that is, the number of photons produced
in our simulation following ~\cite{bib:kakimoto} (FY$_{K}$)
was either increased or
decreased by an arbitrary percentage. In the reconstruction procedure
the original FY$_{K}$~\cite{bib:kakimoto} was used. 

We would like to stress that a shift on the FY is not equivalent to a
shift on the shower energy. When the number of produced photons along the
shower development is changed, absorption effects and trigger
efficiencies must be re-evaluated and the overall result is different
from a simple systematic shift of the shower energy.

As a result of this analysis, the distribution of reconstructed
energies was not only shifted to either larger or smaller energies
but the shape of the EED was also modified.  The mean of
the EED will shift by approximately the same percentage as the FY. 


Figure~\ref{fig:fy} shows the UHECR spectrum convolved with the
fluorescence EED taking FY errors into account.  As can be seen the
flux times the third power of energy shifts significantly.  It shifts
to larger values when the FY error is positive and in the opposite
case, it shifts to lower values when the FY error is negative.  The
GZK cutoff is also smeared but not enough to be absent from the
spectrum.

In order to analyze the effects of a FY systematic error on the
spectrum, we determined the percentage 
excess of events in relation to the theoretical spectrum
convolved with our simulation of the HiRes-II fluorescence telescope.
Figure ~\ref{fig:fyex} shows the excess of events above a given energy
normalized to the total number of events above $10^{19}$ eV. We
considered positive and negative FY systematic error which
corresponds respectively to a significant increase and decrease of
the event flux up to energies around the GZK cutoff.

\begin{figure}[]
\centerline{\includegraphics[width=13cm]{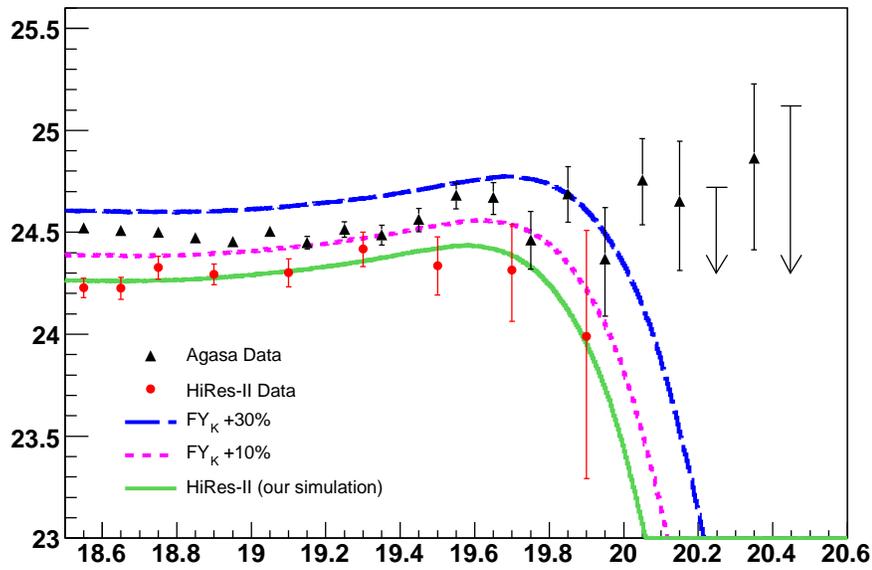}}
\caption{Energy spectrum measured by AGASA and HiRes-II experiments
  compared to a theoretical GZK spectrum convolved with EED
  corresponding to simulations of the fluorescence yield measured by
  Kakimoto et. al and arbitrary shifts of 10\% and 30\%.}
\label{fig:agasa:hires}
\end{figure}

It is clear that an error on the FY will influence the determination
of the GZK cutoff energy. The flux will also be affected
by this error. 
Figure~\ref{fig:agasa:hires} shows the spectra measured by AGASA and
HiRes-II experiments. We also show our calculation of the GZK
theoretical spectrum convolved with the HiRes-II EED. We have
considered three values of the  fluorescence yield in this analysis:
FY$_{K}$ (green solid line), FY$_{K}$+10\% (magenta dotted line) and
FY$_{K}$+30\% (blue dashed line). It can be seen that a FY systematic
error larger than 10\% and smaller than 30\% would be enough to match
HiRes and AGASA 
fluxes but would not smear the GZK cutoff in an important way.

\section{\label{sec:conclusion} Discussion and conclusions}

The influence of various fluorescence telescopes energy error
distributions (EED) in the UHECR energy spectrum was determined
through Monte Carlo simulations.
We first analyzed the intrinsic and unavoidable atmospheric
shower fluctuations. The energy errors that come exclusively from the 
determination of the energy from its longitudinal profile, the 
Gaisser-Hillas fit and missing energy corrections are smaller than 4\%.
We conclude that their effect on the energy spectrum -- when the energy 
is determined through the longitudinal profile -- is negligible.

\begin{figure}[]
\centerline{\includegraphics[width=13cm]{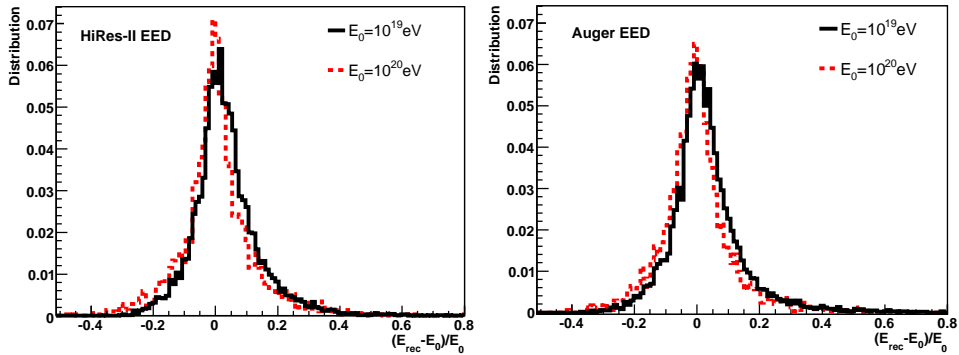}}
\caption{Energy error distribution from simulated fluorescence energy
reconstruction using HiRes-II parameters (left) and Auger parameters
(right).  Two EEDs are shown for each telescope: one using
$10^{19}$~eV  and the other $10^{20}$~eV  proton showers.}
\label{fig:erro:edep}
\end{figure}

\begin{figure}[]
  \centerline{\includegraphics[width=13cm]{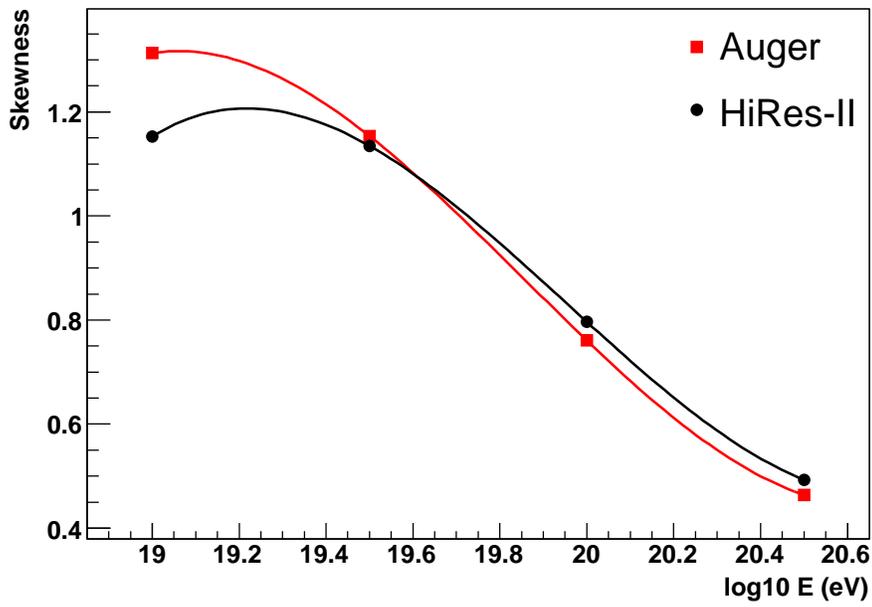}}
  \caption{Skewness parameters for the energy error distribution as a
    function of energy according to our simulations of  the Auger
    (red squares) and HiRes-II (black circles) telescopes.}
\label{fig:skewness}
\end{figure}

Inclusion of detection and energy reconstruction simulation of fluorescence
telescopes results in a considerable
broadening of the EED  (Figure~\ref{fig:erro:edep}). The simulated fluorescence EED
can not be described by a Gaussian nor by a lognormal
distribution. Moreover, the shape of the EEDs change with energy as shown by their
skewness parameter (Figure~\ref{fig:skewness}).

\begin{figure}[]
\centerline{\includegraphics[width=13cm]{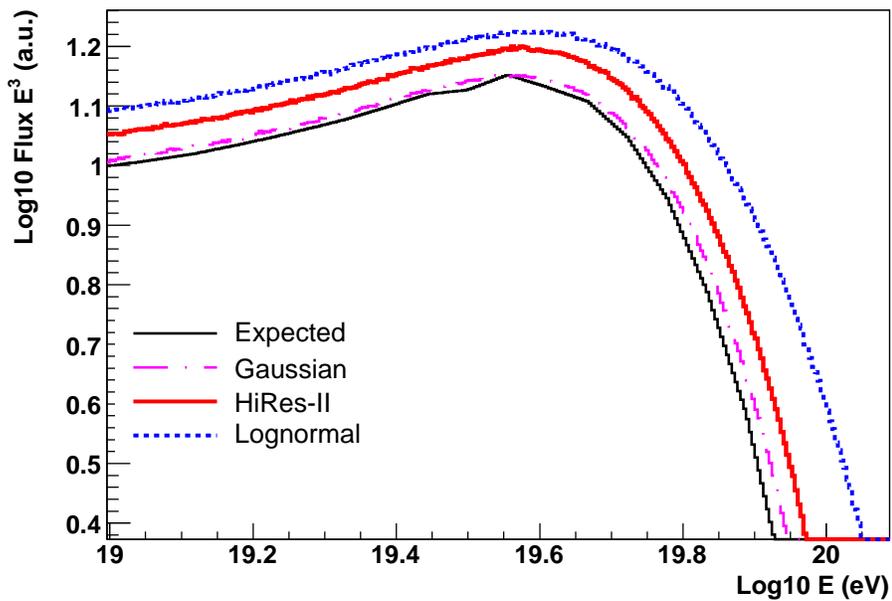}}
\caption{Energy spectrum as expected from theoretical prediction (black 
solid line) and convolved with various energy error distributions. The
red (light) solid curve represents the convolution with the fluorescence EED
determined from our simulation of the HiRes-II telescope; the
magenta dash-dotted line represents a Gaussian convolution with $\sigma =
0.1{\rm E}$ and the blue dashed line a lognormal convolution with
$\sigma(\log_{10}E) = 0.1$}
\label{fig:spectrum}
\end{figure}

\begin{figure}[]
\centerline{\includegraphics[width=13cm]{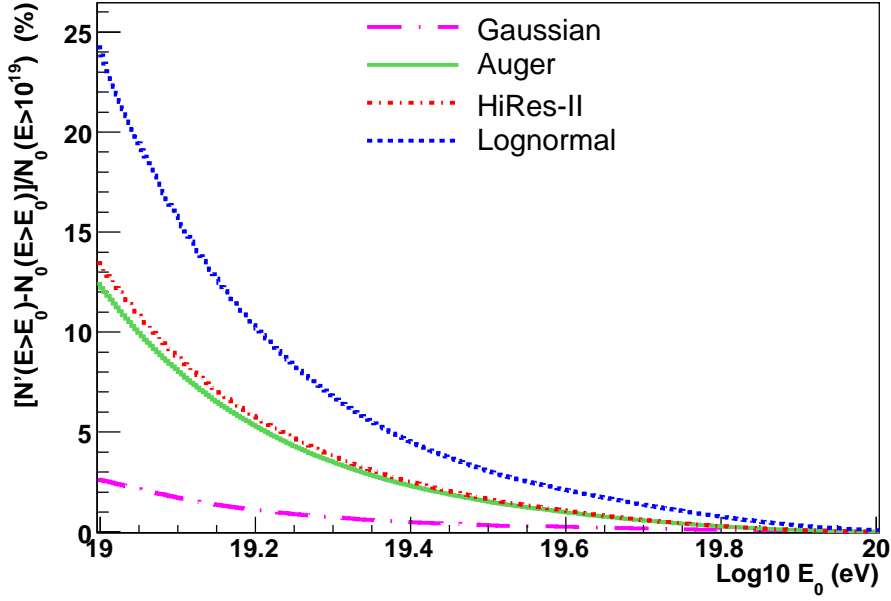}}
\caption{Percentage excess of reconstructed events above
$10^{19}$~eV due to the smearing of the UHECR spectrum with an EED from
our simulation of the HiRes-II and Auger fluorescence telescope and
from a Gaussian with $\sigma = 0.1 E$ and a lognormal with $\sigma =
0.1 $. N' is the number of events above $E_0$ calculated
for each distribution, $\rm{N}_0$ is the number of events above $E_0$
calculated with the theoretical GZK spectrum.}
\label{fig:diff}
\end{figure}

We convolved the UHECR spectrum with fluorescence EEDs
determined from four shower simulations with primary energies of $10^{19}, \
10^{19.5}, \ 10^{20}$ or $10^{20.5}$~eV. These EEDs were generated simulating either
the HiRes-II or the AUGER telescopes. Similar results were obtained for
both telescopes despite the different quality cuts applied. The
convolved spectrum changes in shape and flux as shown in Figure~\ref{fig:spectrum}. 
In order to illustrate the EED's effect on the spectrum around the GZK cutoff we 
determined the excess of events (in percentage) above a given energy. Figure~\ref{fig:diff} 
shows that this effect on the spectrum can result in 5\% more events above
$10^{19.2}$~eV.

\begin{figure}[]
\centerline{\includegraphics[width=13cm]{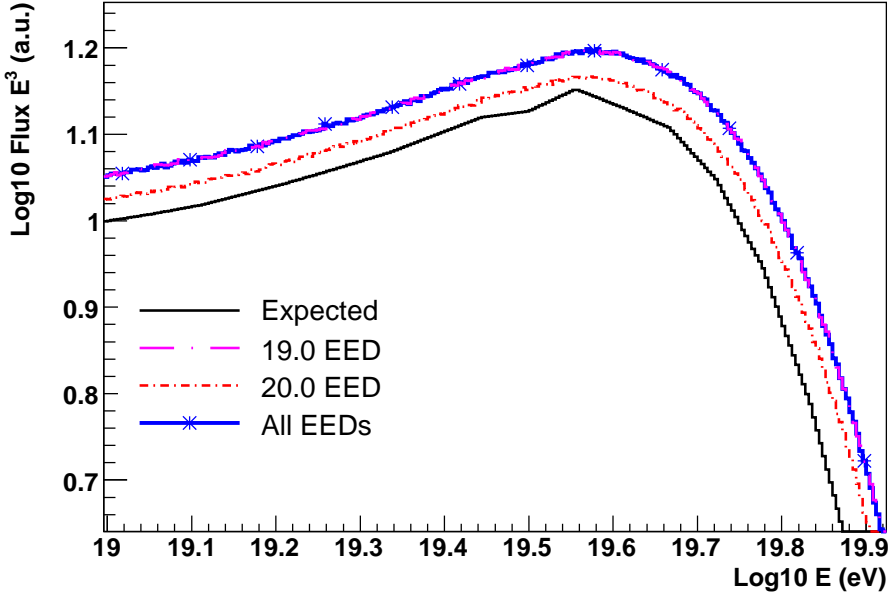}}
\caption{UHECR spectrum convolved with single EEDs (as labeled) and
taking  into account all EEDs. Single EEDs were
determined from $10^{19}$ and $10^{20}$ eV proton showers. The
convolution with the $10^{19}$ EED falls on top of the one with all
EEDs.} 
\label{fig:spec:edep}
\end{figure}

We also investigated the importance of taking the EED energy dependence into
account when convolving the energy spectrum. We conclude that the EED energy
dependence does not need to be taken into account if the EED is determined 
from showers with primary energies that have the largest weight in the flux.
As an example, if the EED is determined from $10^{19}$~eV, the smearing of
the energy spectrum will be the same as if the energy dependence is taken into
account. If however one convolves the spectrum with an $10^{20}$~eV EED, the 
convolved spectrum will be significantly different from the one which is convolved
taking the EED energy dependence into account. This is shown in 
Figure~\ref{fig:spec:edep}.  

One of the most fundamental parameters in the energy reconstruction is
the fluorescence yield (FY). We have analyzed the influence of a
systematic error in the FY on the energy spectrum. The production of
fluorescence photons in our simulation was increased by an arbitrary
factor while the FY in the reconstruction procedure was always fixed to the
value determined by Kakimoto et al.~\cite{bib:kakimoto}. 
As stressed in
section \ref{sec:fy}, shifting the FY is not equivalent to an automatic
shift in the reconstructed energy. Not only the average reconstructed energy 
shifts systematically by the same FY error factor as the EED has its
shape modified. The smearing of the energy spectrum
due to the FY errors is greatly enhanced when compared to the spectrum 
convolution with no FY errors.

\begin{figure}[]
\centerline{\includegraphics[width=13cm]{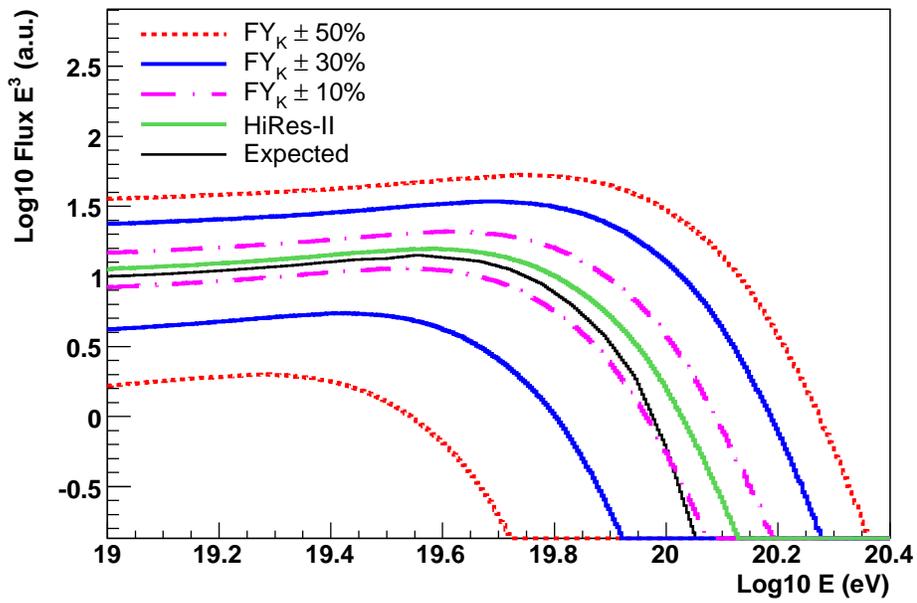}}
\caption{UHECR spectrum (dark solid line) and convolved with an EED
from our simulation of the HiRes-II fluorescence telescope with no FY
systematic errors. Convolution with EEDs from
simulations including systematic errors are labeled.}
\label{fig:fy}
\end{figure}

Figure~\ref{fig:fy} shows the GZK spectrum convolved with the HiRes-II
simulation which included FY errors of $\pm$10\%, $\pm$30\% and $\pm$50\%.
Spectrum convolutions with fluorescence EEDs that include positive FY errors are
not symmetric in relation to the ones with negative FY errors.
It is clear that the modulation of the
spectrum is not symmetric from positive to negative FY errors.

\begin{figure}[]
\centerline{\includegraphics[width=13cm]{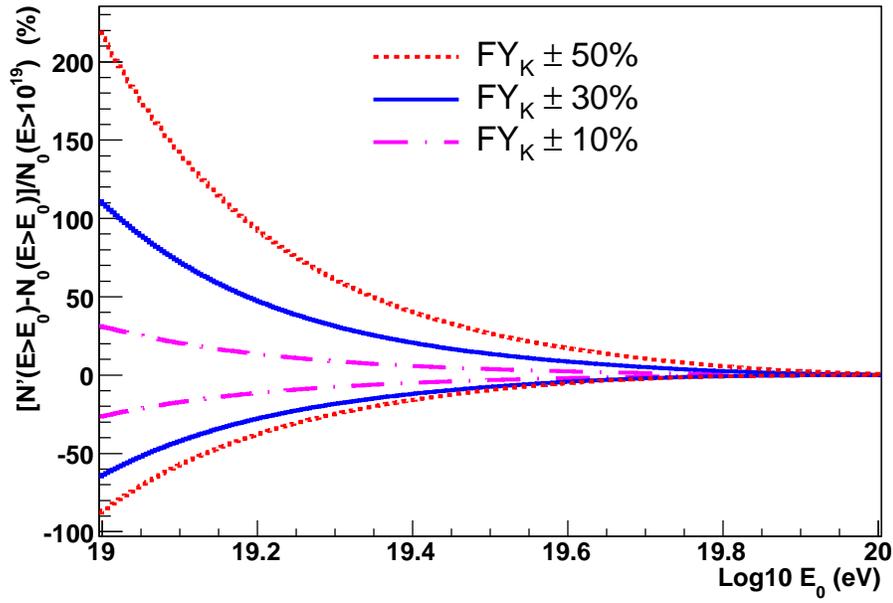}}
\caption{Percentage excess of reconstructed events above $10^{19}$eV
due to the smearing of the UHECR spectrum with an EED determined from
our simulation of the HiRes-II fluorescence telescope including FY
systematic errors. N' is the number of events above $E_0$ calculated
for each FY error case, $\rm{N}_0$ is the number of events above $E_0$
calculated with the GZK spectrum convolved with the HiRes-II energy
error.  There are positive and negative percentages corresponding
respectively to positive and negative FY systematic errors.}
\label{fig:fyex}
\end{figure}

We conclude that taking the FY errors into account moves the GZK significantly.
The position of the GZK cutoff in the spectrum might be shifted to higher or
lower energies depending on the sign of the FY error. Figure~\ref{fig:fyex}
shows that this error can result in significant increase or
decrease of events depending on it's sign. We also conclude that
although the GZK cutoff position might shift significantly it will
not be erased.

The measured flux is also directly proportional to the FY error. A
error larger than 10\% and smaller than 30\% of the FY is enough to match the
flux measured by the HiRes and the AGASA collaborations.

Finally, we conclude that the energy error distributions of
fluorescence telescopes including shower fluctuations, detection and reconstruction
uncertainties and fluorescence yield errors will significantly smear the 
UHECR energy spectrum. The GZK cutoff position in the spectrum might shift
significantly but not enough to erase the GZK cutoff.

{\em Acknowledgements --}
We thank Bruce Dawson for his comments. W.C. and V.S. are supported by
the State of S\~{a}o Paulo Research Foundation (FAPESP).

\bibliographystyle{elsart-num}

\bibliography{allbib.bib}

\newpage

\begin{table}
\begin{center}
\begin{tabular}{|l|c|c|} \hline \hline 
CUT & HiRes-II & Auger \\ \hline 
Angular Speed & $<\; 11^{o}/\mu s$ & \\ 
Triggered PMTs & $>\; 7$ & $>\; 5$ \\ 
Track Length ($< \;17^o$ elevation) & $> \; 7^o$ & \\ 
Track Length ($> \;17^o$ elevation) & $> \; 10^o$ & \\ 
Path Length & & $>\; 200$ g/cm$^2$ \\
Zenith Angle & $< \; 60^o$ & $< \; 60^o$ \\ 
Minimum Viewing Angle & & $> \; 18^o$ \\ 
$X_{max}$ & visible & visible \\ 
$\chi^2/{\rm d.o.f}$ (GaisserHillas fit) & $<\; 10 $ & $<\; 20 $ \\  \hline \hline
\end{tabular}
\caption{Quality cuts applied in the UHECR energy spectrum analysis by
Auger \cite{bib:auger:espectro} and HiRes \cite{bib:hires:espectro}
and included in our simulations.}
\label{tab:cuts}
\end{center}
\end{table}

\begin{table}
\begin{center}
\begin{tabular}{|c|c|c|c|c|} \hline \hline 
$Log10 E (eV)$ &\multicolumn{2}{|c|}{Auger} &\multicolumn{2}{|c|}{HiRes-II} \\ \hline      
$Log10 E (eV)$ &Mean &RMS  &Mean &RMS \\ \hline 
19.0 & 2.87\% & 11.67\% & 3.38\% & 11.01\%  \\ \hline 
19.5 & 2.23\% & 12.11\%  & 3.22\% & 12.13\%  \\ \hline 
20.0 & -0.30\% & 11.43\% & 0.75\% & 12.10\%  \\ \hline 
20.5 & -3.30\% & 9.98\% &-2.30\% & 10.14\%  \\    \hline \hline
\end{tabular}
\caption{Average and RMS of the energy error distributions as a
function of  energy for our simulation of the HiRes-II and Auger
Telescopes.}
\label{tab:en:erro}
\end{center}  
\end{table}

\end{document}